# Just In Time Indexing

Pinaki Mitra

Indian Institute Of Technology, Guwahati

pinaki@iitg.ernet.in

Girish Sundaram

India Software Lab, gisundar@in.ibm.com

Sreedish PS

Indian Institute Of Technology, Guwahati

sreedish@iitg.ernet.in

*Abstract*—One of the major challenges being faced by Database managers today is to manage the performance of complex SQL queries which are dynamic in nature. Since it is not possible to tune each and every query because of its dynamic nature, there is a definite possibility that these queries may cause serious database performance issues if left alone. Conventional indexes are useful only for those queries which are frequently executed or those columns which are frequently joined in SQL queries. This proposal is regarding a method, a query optimiser for optimising database queries in a database management system. Just In Time(JIT) indexes are On Demand, temporary indexes created on the fly based on current needs so that they would be able to satisfy any kind of queries. JIT indexes are created only when the configured threshold values for resource consumption are exceeded for a query. JIT indexes will be stored in a temporary basis and will get replaced by new JIT indexes in course of time. The proposal is substantiated with the help of experimental programs and with various test cases. The idea of parallel programming is also brought into picture as it can be effectively used in a multiprocessor system. Multiple threads are employed while one set of threads proceed in the conventional way and the other set of threads proceed with the proposed way. A live switch over is made when a suitable stage is reached and from then onwards the proposed method will only come into picture.

## I. INTRODUCTION

Database systems have evolved over a period of time from simple and homogeneous to more complex and diversified. Database Queries also have become more and more complex and resource intensive because of business needs. It is common knowledge that one of the major challenges being faced by Data Base Administrators today is to manage the performance of complex SQL queries which are dynamic in nature. There could be multiple users using the system and each end user requirement may vary significantly and so would be the dynamic query generated to fetch data from the database. These queries may be requiring many joins and sorts which can eventually lead to the slowing down on of the whole system because of many disk reads. Conventional indexes are useful only for those queries which are frequently executed or those columns which are frequently joined in SQL queries. There can be the case that there is no suitable index which can meet the requirement of a particular query which may be is going to make some heavy joins or sorts. One way to improve performance of queries is to ensure that there are proper indexes available. Indexes are generally created on table columns which are frequently joined in queries and for those queries which are executed frequently. There are options for the database administrators to manually create indexes. It is not a feasible option to monitor each and every querys execution plan and to identify whether the query is going to use any index or it is going to fetch data directly from the table. The biggest challenge for building the temporary indexes is the trade off between the building cost of a temporary index and the worthiness of it. In multiprocessor systems, a new process can be spawned for building the temporary index while the old process will start digging the database and once the temporary index is ready a live switch over to the index can be made. But all these things heavily depend on the estimation of query resource consumption and the

Pinaki Mitra is with the Department of Computer Science and Engineering, Indian Institute of Technology, Guwahati email: pinaki@iitg.ernet.in

Girish Sundaram is with IBM India Software Labs email: gisundar@in.ibm.com

Sreedish P S is with the Department of Computer Science and Engineering, Indian Institute of Technology, Guwahati email: sreedish@iitg.ernet.in

trade off between making cost of new index and the cost of answering queries without any index.

A. Motivation

The Phantom queries which are executed only once in a while or just once based on specific user needs are the real threat for the database systems which can even hang up the entire system for hours. In many situations the user cannot even predict when the output will come or even whether the output will come or not. A reasonable amount of uncertainty prevails in many queries. More often than not there will not be any indexes on the table which may be used to improve the performance of these queries. Since it is not possible to tune each and every query because of its dynamic nature, there is a definite possibility that these queries may cause serious database performance issues if left alone. The leading database vendors have the mechanism to monitor online query performance by using a variety of tools both graphical and command line. For e.g.

1) Oracle has a tool called as the Performance Manager used for analyzing online performance of database taking into consideration areas like disk I/O, contention, memory etc. Another tool commonly used is TOP Sql used to find out queries which are the most resource intensive.
2) DB2 has tools like the Query Patroller used for monitoring online performance of queries and alert the DBA if any query uses more resources than that specified by the threshold.

There are many drawbacks for these existing solutions such as

1) Although they provide comprehensive information about the queries taking the most resources but they do not take any steps to find out the root cause why these queries are taking more time to execute.
2) Most often it is left to the DBA to either terminate the queries manually or set the threshold to run these queries at a lower priority. Hence the problem is not solved but only avoided.
3) None of the existing solutions provide any suggestions or recommendations to avoid the problem in the future.
4) The DBA may actually go ahead and terminate the SQL but it may actually be an important process and may need to be run again. There is a definite possibility that the problem may happen again when the query is re-executed.

Hence there is a need to have a new solution which will not only prevent the occurrences of run time queries hogging the system due to lack of proper indexes, but also take steps to ensure that they are executed after rectifying the problem area.

B. Objective

The proposal is related to the database management field, more specifically it is related to a method, query optimiser and a computer program product for processing and optimising database queries in a database management system. Just In Time Indexing is a proposal for a temporary indexing technique which will create on the fly and dynamically based on the incoming query. The main objective of this project is to substantiate the proposal of just in time indexing through various theoretical aspects. A test environment is to be developed and the experiments are to be conducted for getting a practical support for the proposal. Various parameters are to be fixed during the various stages of just in time indexing.

C. Problem Definition

The decision made on whether to create the temporary index cannot be always be accurate or provide the greatest efficiency. The correct decision on one time may not be the correct decision at another time because it will depend on many factors such as the system workload, whether the files or indexes happen to be in memory or not etc[5]. Our aim is to create an automated system which will create temporary indexes on the fly as shown in the example and dynamically based on the query. Various issues are to be addressed like when and how to create the index. Another important challenge is to create indexes intelligently so that the indexes created can really be useful rather than indexes which are wasting resources.

## II. EXISTING QUERY OPTIMISING TECHNIQUES

A database management system in general includes a query optimiser that attempts to optimise the performance of a query after making the query execution plan. After the query compilation and before the query execution the optimiser generally tries to optimise the resource consumption of the query. The query optimiser selects from multiple query plans possible or possible implementations of a query to execute the query with the greatest efficiency[8]. The first part of the survey was to look on to the various query performance monitoring systems used by the leading database vendors. As a part of that, the query performance monitors of Oracle, DB2 and Apache Derby were investigated. These leading database vendors have the mechanism to monitor online query performance by using a variety of tools both graphical and command line.

For e.g.
1) Oracle has a tool called as the Performance Manager used for analysing online performance of database taking into consideration areas like disk I/O, contention, memory etc.. Another tool commonly used is TOP SQL used to find out queries which are the most resource intensive.
2) DB2 has tools like the Query Patroller used for monitoring online performance of queries and alert the DBA if any query uses more resources than that specified by the threshold.
3) Apache Derby has tools like XPlain which is capable of generating XML reports on query resource consumption. It can be configured even to build tables which holds data about the various query resource consumption as tuples on which we can later query on to get the resource consumption details of a particular query.

### A. Drawbacks of these existing solutions
1) Although they provide comprehensive information about the queries taking the most resources but they do not take any steps to find out the root cause why these queries are taking more time to execute.
2) Most often it is left to the DBA to either terminate the queries manually or set the threshold to run these queries at a lower priority. Hence the problem is not solved but only avoided.
3) None of the existing solutions provide any suggestions or recommendations to avoid the problem in the future.
4) The DBA may actually go ahead and terminate the SQL but it may actually be an important process and may need to be run again. There is a definite possibility that the problem may happen again when the query is re-executed.

### B. Observations

Hence there is a need to have a new solution which will not only prevent the occurrences of run time queries hogging the system due to lack of proper indexes, but also take steps to ensure that they are executed after rectifying the problem area. All these query optimisers selects from multiple query plans of the query to achieve the greatest efficiency[8]. The query optimisers are faced with a dilemma with regarding query execution is whether to use an existing index, or to build a new temporary sparse index or to work as if there is no index. This dilemma exists for queries that have ordering or grouping criteria from one table. The trade off between the choices of using an existing index or a temporary sparse index can be looked up as follows. An existing index has no start up time to use it since it is already built, but using that index may result in long fetches because of extra I/Os on a table object typically must be performed to distinguish between the records that are to be discarded and that are to be returned to the user. On the other hand, the temporary index has a start up cost to build it, but once built, the fetch up time will be much faster if the Index is made wisely. The fetch time is faster for sparse index because every record processed in the index will be one of interest and no time is wasted for unnecessary records.

## III. Techniques Used in Just In Time Indexing

The proposed algorithm for JIT Indexing framework consists of three main modules. Before going to the complete proposed algorithm the individual components are explained in this chapter. Along with the description of individual modules, the challenges in designing the modules are also discussed thoroughly in this chapter.

### A. JIT Alert Process

This process jump starts the JIT indexing engine to life once the resource consumption of the user query crosses the defined threshold. This module is responsible for monitoring the query execution plan made by the Database manager. This module has to check whether the resource consumption of a incoming query is exceeding some threshold. The term resource consumption includes various parameters such as number of row fetches, whether any sorting is needed or not, the time that is going be consumed for answering the query etc. The threshold can be set according to the size of the database or according to the user convenience. The setting of threshold plays a crucial role in activating the whole Just In Time Indexing process. The threshold has to be set sensibly for the best results out of Just In Time Indexing. Based on these observations and some normalisations JIT Alert process will come up with a integer value and that value will be matched with the defined threshold. If the normalised resource consumption cross the threshold, the JIT indexing engine will get activated. JIT Alert process is also responsible for the live monitoring of query resource consumption. In most of the cases prediction of the resource consumption by the query is very tedious. So the better approach is to make a live monitoring of the resource consumption of the query. Whenever the query resource consumption is crossed, JIT Alert process also needs to calculate the amount of data fetches remaining for answering the query. Based on all these JIT Alert process has to decide whether to trigger the whole process or not.

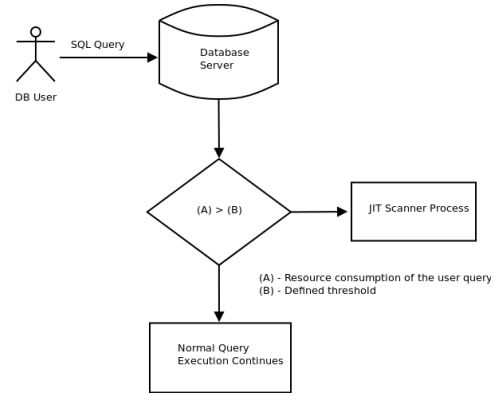

Fig. 1. JIT Alert Process

### B. JIT Scanner Process

This process is responsible for scanning the incoming user SQL queries and finding out whether existing indexes can be used to satisfy the query. This module will keep track of all the existing indexes and their detailed information. Based on this information this module decides whether to create a new index or to use an existing one. If an existing suitable index is found, the JIT engine will exit and the query will use that index. This module has to go hand in hand with the database manager because only from the query optimisers of the database manager this module can fetch information whether the query is going to be answered by a table scan or by a index scan. JIT Scanner process will also keep information about all the JIT indexes created before in a database and whenever the query optimiser of database manager answers that an index scan is going to be performed, JIT Scanner process can conclude whether its a JIT index or not.

### C. JIT Indexer Process

This process is responsible or creating the JIT index for the specific query and maintaining it. If the JIT Scanner Process initiates the creation of a new index, this module makes calculations for the cost of a non indexed query reply and cost of indexed query reply. If cost indexed query reply is less than that of an non indexed query reply, this module builds a new customised index which is suitable for the query and answers the query with the index. Index creation and maintenance are also maintained by this module. This module also tracks the information about the indexes currently maintained and the time when the indexes are used last. It is a kind of refresh counter and whenever a new Just In Time Index is created and the number of total indexes are above a particular limit, JIT Indexer process pops the index which is used for the least and updates the open ended stack with the latest index. Thus this module prevents the system getting clogged because of too many indexes.

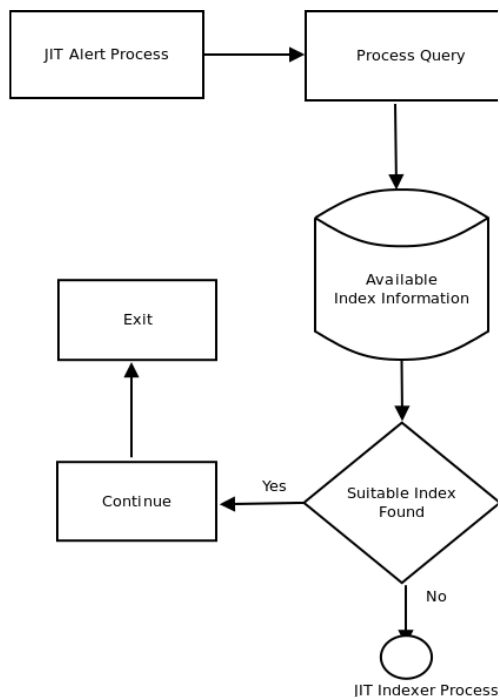

Fig. 2. JIT Scanner Process

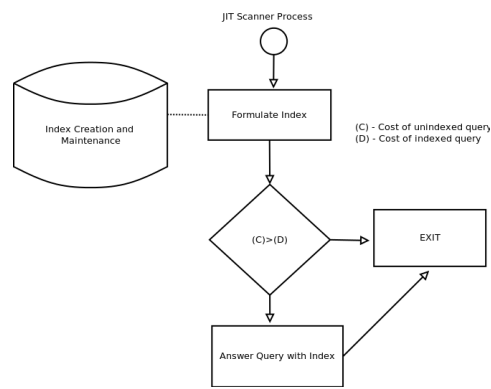

Fig. 3. JIT Indexer Process

### D. Challenges with JIT Alert Process

The main goals to be met in this module are to estimate the query resource consumption and to define the threshold for which the JIT engine should be activated.

*1) Estimation of query resource consumption:* Whenever a query comes into the database server, it is first parsed and checked for errors. Then the database manager usually comes up with multiple query execution plans and the best query execution plan with the minimum resource consumption is selected for actual execution of the plan. Here in JIT way of answering queries, queries have to pass a test by JIT indexing engine. The query resource consumption has to estimated without actually executing the query by various methods. Application of heuristics or meta data about database tables can be used for forecasting the resource consumption of the query. The joins present in the query play a vital role in the query resource estimation since the order in which joins are evaluated is very important.

*2) Estimation of threshold:* The definition for a good threshold is also very much important for triggering the whole process. Whenever the query resource consumption crosses the defined threshold, the JIT indexes should come into play. The consideration of all the factors used in estimating the threshold is crucial in determining the threshold. Threshold can be made as a static one or as a dynamic one. A dynamic one will be more meaningful as it can server like a self learning system. The average query resource consumption of all the queries ran so far can be used as a good threshold. Storing additional information about the queries executed so far is a trivial requirement if the threshold estimation is made dynamic.

### E. Challenges with JIT Scanner Process

This module is responsible for deciding whether any existing conventional index or previously made and yet existing Just In Time index cane be made useful for this process. This module has to maintain information about all the indexes kept in hand. This module has to keep track of outdated indexes also. Whenever an index is swapped out, this module has to update its database. So the available index data maintenance and giving informations about it is the main challenge for this module. This module also need to look up whether the database manager is going to answer the query using a table scan or using an index scan. If it is a table scan, this module has to check whether any suitable Just In Time index is there or not. Whenever a suitable index is found for the incoming query this module should make use of that index and if there is no suitable index this module should inform the JIT Indexer Process that there is no suitable index for this heavily resource consuming query.

### F. Challenges with JIT Indexer Process

This module is in charge of creating the indexes which are of to be useful with the incoming query. This module has to analyse the query for the joins and for the 'where' clause and according to that this module has to make indexes which will turn out to be useful index. If this module comes out with a wrong plan for an index, the whole effort will be in vain and that emphasises the importance of this module. Making of useful indexes for the incoming query and for the future queries are the main challenge for this module. The comparison of cost of un-indexed query reply and indexed query reply needs to calculate the resource consumptions and these also come under the challenging part of JIT Indexer process.

## IV. INDEX CREATION STRATEGY

The index creation strategy used by JIT Indexer Process is to analyse the incoming query and then to formulate an index accordingly. Data Manipulation Language (DML) statements are used for managing data within schema objects. A DML 'select' statement will be having a 'where' clause where it specifies the evaluation criteria. Accordingly the JIT Indexer will formulate the index creation.

For example, If the query is like 'SELECT * from T WHERE ROWNUM $<=$ 10', the the JIT indexer finds that the index that can be of use to this particular query is ROWNUM. So an index on the column ROWNUM will serve the purpose for this query.

This simplicity will not work if the 'where' clause specifies more than one column for its evaluation criteria using some logical operators. For example, Suppose the query is like 'SELECT * from T WHERE ROWNUM $<=$ 10 and WEIGHT $>$ 5'. Two columns are involved in this query a total of three indexing possibilities are there.

1) An index on column ROWNUM can alone be created
2) An index on column WEIGHT alone can be created
3) A Multi index on both ROWNUM and WEIGHT can be created

To generalise, if there are N columns involved in a query, the total indexing options possible are

$$\binom{N}{1} + \binom{N}{2} + \cdots \binom{N}{N}$$

which is asymptotically equal to $2^N$. Out of which N are single column indexes and all others are multi column indexes.

### A. Index Search Problem- A Search Problem having Exponential Time Complexity

Index Search Problem is proved to be NP-hard by [3]. It can be viewed as a combinatorial optimization problem since its solution has to be chosen from among a finite number of possible configurations. Therefore, explicit and complete enumeration of all the possible index subsets is a possible way to solve Index Selection Problem. For small N, this enumerative approach is probably the best strategy, as the optimal combination of domains to be indexed is guaranteed to be found. This method is however impractical in most cases, since the computational effort grows exponentially with the number of candidate indexes for selection.

## B. Existing literature on Index Selection Problem

Both heuristic and exact approaches for index selection in a relational DB environment have been proposed in the literature. For example [2] [6] [7] [11]. Heuristic algorithms give an approximate solution to the problem, in the sense that they do not guarantee that the index subset they select is the one which minimizes the estimated execution cost of the DB workload. The exact approaches so far proposed in the literature suffer from the fact that they consist of complete enumeration of the possible index subsets, hence are very time consuming even for small problem instances.

## C. Index Creation in Leading Database Manager Programs

Leading database manager vendors and researchers are still relying on heuristics based calculations for finding the best suitable index for a particular query. No database manager has a completely automated index creation mechanism. The database managers always leave the option of creating indexes to the database administrator along with some suggestions for the best index. The database administrators usually will be having a reasonable amount of information about the queries which are coming ahead and accordingly they manually create indexes.

## V. INDEX CREATION TECHNIQUE USED IN JUST IN TIME INDEXING

In Just In Time Indexing, the index creation is completely automated and based on heuristics. As a consequence there can be some instances where the automation can fail for coming up with the best index. Just In Time Indexing consider many statistical facts about the instance of the database at a particular point of time and meta data for coming up with a suitable index. The idea is to come up with multiple indexing options based on heuristics and within them find the best index. Even though it might not be the best index, it can be assured that it is not the worst index.

## A. Index Selection Heuristics

1) Some attributes can be eliminated from the initial formation of candidate set of indexes by the virtue of their low occurrence frequencies in queries. By gathering statistical data on the incoming queries we can know which columns are used and joined more for answering user queries in a database management system. According to this we can rank the attributes by their frequency of usage and candidate indexes formed by low rank domains can be omitted. An upper bound P can be put on the number of domains that should take part in the making of candidate indexes.

2) Every index will be associated with two values. The number of page accesses that an index on the column can save on the processing of the query and a estimated maintenance cost of the index. If the upper bound on the number of page accesses saved by an index is less than the projected cost of maintaining an index, that column can be safely excluded from forming the candidate set of indexes. Using the selectivity estimate of the domain, we can find a gross upper bound on the number of page accesses that an index on the domain can save in the processing of the forecasted set of queries[9]. If this upper bound is less than the projected cost of maintaining an index on the domain, then this domain can safely be excluded from the initial candidate set.

3) It is always good to come up with indexes for columns having relatively unique values.

4) We can rank the domains using the above heuristics and can consider only the top ranking M domains. Only these M domains and combinations of these domains are used for formulating the candidate indexes for the best index. Apart from this, a bound M' which is less than M can be also put on the number of domains that is to be considered together for forming multi indexes.

## VI. HEURISTICS BASED ALGORITHM FOR FINDING THE BEST INDEX

Let D={$d_1$, $d_2$, ...$d_i$} be the domains that has to be accessed by Query Q.

Let J={$j_1$, $j_2$, ...$j_k$} be the resulting index permutations such that the upper limit of number of domains included in a particular index is M'.

1: for each $d_i$ D do
2:    sort in ascending order in accordance with the frequency of previous occurrence
3: end for
4: Select first P domains from the sorted list of domains and reset D using these P domains
5: for each $j_i$ J do
6:    Using the heuristic 2 sort the candidate indexes based on their maintenance cost versus page accesses saved ratio
7: end for
8: Select top T' number of candidate indexes from the sorted list and make the set T
9: for each $t_i$ T do
10:    Create hypothetical indexes using the STATISTICS_ONLY option of the CREATE INDEX statement
11: end for
12: Find the best index from the T' indexes using the recommendation from the query optimiser using EXPLAIN statements and actually create the index without the STATISTICS_ONLY option.
13: if Tie in Step 12 then
14:    break tie using the index space consumption
15: end if
16: Drop all the remaining hypothetical indexes.
17: EXIT

### A. Algorithm Explanation

Thus Just In Time Indexer will come with multiple indexing plans in accordance with the heuristics. Algorithm is self explanatory except for hypothetical indexes. Creating real indexes in a best index finding session is not feasible because its overhead could impact operational queries and degrade the performance of the whole database. So the challenge is to estimate the cost of using an index that does not yet exist. So, to avoid creating indexes during a best index finding session, JIT engine uses a special kind of indexes called hypothetical indexes. As the name implies, hypothetical indexes are not real indexes; they only contain statistics and can be created with the undocumented WITH STATISTICS_ONLY option of the CREATE INDEX statement which is part of the standard query language. With the help of hypothetical indexes, the database engine's query optimiser will give recommendations for which index to use. Accordingly JIT indexing process initiates the actual creation of the recommended index and to drop the hypothetical index. The database can use indexes more effectively when it has statistical information about the tables involved in the queries. JIT gathers statistics when the indexes are created by including the keywords COMPUTE STATISTICS in the CREATE INDEX statement.

### B. Algorithm Evaluation

This heuristics based algorithm can work fairly well as it is concentrating on the factors which are crucial for an index. Live statistics gathering will enhance the adaptation towards the dynamic and continuous changing of database instances. Once an optimal set of domains for indexing has been chosen, the total cost for this choice of domains, including query processing cost, index storage and maintenance cost are compared in between each other and the best ones evolve after every filtering stage. The final index may not be the optimal one as the whole thing is based on certain heuristics. There can be the case like the optimal domain for the index might have ruled out in the first filtering step itself. Suppose if the column is referred for the first time and it is part of the optimal solution. The column will be ruled out in the first step itself as the frequency of use of that particular domain in answering any query is zero. But on most of the cases the winner will the real optimal index or very near to optimal index.

## VII. ALGORITHM FOR JUST IN TIME INDEXING

One experimental algorithm is proposed for the same. This algorithm will be added with more features taking into account about the availability of multi processor systems.

## VIII. ALGORITHM FOR JUST IN TIME INDEXING

1: User submits query for database processing
2: JIT_Alert process compares the resource consumption of the User query (A) versus the defined threshold (B).
3:   if (A) > (B) then
4:     goto step 8
5:   else
6:     continue with step 2
7:   end if
8: JIT_Scanner process scans the existing indexes which can be used by the User query
9:   if match found then
10:    goto step 21
11:  else
12:    goto step 14
13:  end if
14: JIT_Indexer process formulates the best possible index making use of cost optimization for the user query
15: JIT_Indexer process compares the cost of un-indexed query (C) versus indexed query (D)
16:  if (C) > (D) then
17:    goto step 21
18:  else
19:    goto step 23
20:  end if
21: JIT_Indexer process creates the JIT index for the query and stores the index data in the temporary memory space (open ended stack)
22: JIT_Indexer process deletes index and de-allocates JIT Indexes which are not being used by using the standard ageing mechanism.
23: EXIT

### A. Algorithm Description

Just In Time Indexing algorithm gets initiated with the user submitting query for database processing. The first module is JIT Alert Process and this module calculates the resource consumption of the user query with the defined threshold. If the resource consumption of the user query is less than the defined threshold then nothing has to be done. If the query resource consumption is crossing the defined threshold, then the JIT Alert process initiates the next module which is JIT Scanner process. The role of JIT Scanner process is to scan all the existing indexes which can be used for answering the user query. If a matching index is done then the task is over for the JIT process and all that is remaining is to answer the particular query with the found index. But if there is no suitable index in the custody of JIT Scanner process, then JIT Scanner process initiates JIT Indexer process. JIT Indexer process formulates the best possible index for making use of the cost optimization for the user query. After formulating the index, JIT Indexer process compares the cost of un-indexed query answering versus indexed query answering. If indexed query answering is found to be better effective, then the query processing is proceeded with the formulated index. The formulated index is kept in a temporary memory space and the old indexes are deleted according to some standard ageing mechanism. The aged out indexes are deleted from memory in time to time to prevent clogging of the whole system. The pictorial representation of the whole JIT Indexing process is demonstrated in figure 4.1

## IX. EXPERIMENTS AND RESULTS

### A. Experimental Setup

In this section the softwares selected for the experimental purposes are discussed. The questions like why this particular software is addressed in this particular section.

*1) Apache Derby:* The Apache Derby project is a relational database implemented entirely in Java. It's key advantage is that it is based on the standard Java JDBC and SQL standards. It is also easy to install, deploy and use as well as it can be embedded in almost any light-weight Java application. Derby provides an embedded JDBC driver that lets you embed Derby in any Java-based solution. The database manager Apache Derby was selected for experimental purposes. Apache Derby is a relational database manager which can be easily embedded into programs. This choice was made mainly because Apache Derby is an open source database manager. One can view the code of internals of Apache Derby without much effort since it is open source. Things would have been more difficult with other proprietary databases. Moreover the signature of Apache Derby is very small which is nearly 2 MB.

2) Java Database Connectivity(JDBC): This technology is an API for the Java programming language that defines how a client may access a database. It provides methods for querying and updating data in a database. JDBC is oriented towards relational databases.Also it is found out that Java Database Connectivity (JDBC) is very handy for database connectivity. It provides methods for querying and updating data in a database. JDBC is oriented towards relational databases. JDBC allows multiple implementations to exist and be used by the same application. The API provides a mechanism for dynamically loading the correct Java packages and registering them with the JDBC Driver Manager. The Driver Manager is used as a connection factory for creating JDBC connections. JDBC connections support creating and executing statements. These may be update statements such as SQL's CREATE, INSERT, UPDATE and DELETE, or they may be query statements such as SELECT. Additionally, stored procedures may be invoked through a JDBC connection.[14].

3) Java: Java is a general-purpose, concurrent, class-based, object-oriented language that is specifically designed to have as few implementation dependencies as possible. It is intended to let application developers write code that runs on one platform does not need to be recompiled to run on another. Java is currently one of the most popular programming languages in use, particularly for client-server web applications. For coding purposes of this project the front end was decided to be coded in Java. Swing is the primary Java GUI widget tool kit. Swing provides a native look and feel that emulates the look and feel of several platforms, and also supports a pluggable look and feel that allows applications to have a look and feel unrelated to the underlying platform. In addition to familiar components such as buttons, check box and labels, Swing provides several advanced components such as tabbed panel, scroll panes, trees, tables and lists. The Java swing libraries are excellent for making any window based graphical front end.

4) Xplain Tables: XPLAIN style is an enhanced form of RUNTIMESTATISTICS processing which preserves captured statistics information in database tables. Once the statistics have been collected and saved in the tables, they can be queried for analysis purposes. The current XPLAIN implementation is optimized for ad-hoc queries and tool support. Furthermore, the explain data is quite extensive to analyse. Derby tries to implement a compromise between detailed explain information which is almost unreadable by human users and which has to be evaluated with the help of a tool, versus a compact version of explain data which is only applicable for rough investigations but is still browseable by human users. We feel that the information in the XPLAIN system tables is sufficiently detailed to be powerful, but still simple enough to provide useful information to ad-hoc querying during interactive use. This Just In Time Indexing implementation seeks extensive help from the Xplain Tables of Apache Derby.

B. Experimental Results

Java Swing front end, JDBC and Apache Derby are the core utilities used for making this proposal. A sample database was made for testing purposes. It is a very big database consisting of five tables, each having about one million tuples and seventy attributes. For a better understanding of the database, one real scenario was simulated for making the database. This database simulates the marks scored by one million students in an exam consisting of five subject tests. Each subject exam had sixty five questions and each table which corresponds to a subject exam stores the student's unique id, students name, marks of sixty five questions in sixty five columns, total marks and rank in that subject. The tables are populated with data using scripts which generate random data. Total marks and rank were calculated using procedures.

One window based front end was made JIT indexing for experimental purposes. When the application is loaded, it automatically establishes connection with the above said database. This front end has a text area through which we can fire queries into the exam database. When a query is fired through this application it will be in compile only mode. ie The queries will be compiled by the database manager but they will not get executed. By compiling it means that the database manager will be coming up with a query execution plan. By investigating in this query execution plan the JIT Alert process can calculate the resource that is going to be consumed by the query. Also it can know whether any existing index is going to be used or not. This informations are gathered using various facilities of Apache Derby. There is provision for making the database queries in "compile-only" mode, which means the queries will be compiled and they will not get executed. By compiling it means that the database manager will be coming up with the best query execution plan. There will be multiple ways for executing a complex query, but the database manager will select the best available plan for executing the plan. Once the query is fired into the database, the query resource consumption is calculated and the query is executed with and without Just In Time Indexing. Separate windows will give information regarding the resource consumed by the query in the normal mode of operation and with Just In Time Indexing. So the comparison can be made very easily. Through this experimental setup abundant amount of queries are fired into the sample database and results were monitored. The sample queries are fired to the database through the software. The database includes five tables, each representing the marks scored by one million students in sixty five questions. Two sample tables, Table I and Table II are demonstrated for visualisation. The tables symbolically representing the marks for chemistry and physics are used for our resource consumption and indexing tests. The contents of the tables are filled using random values and scripts were used for generating the random values and for inserting into the tables.

| P_ID | First_Name | m1 | m2 | m3...m65 |
|---|---|---|---|---|
| 1 | abc | 1 | 3 | ... |
| 2 | acb | 0 | 2 | ... |
| 3 | bca | 3 | 2 | ... |

TABLE I
SAMPLE TABLE FOR PHYSICSMARKS

| P_ID | First_Name | m1 | m2 | m3...m65 |
|---|---|---|---|---|
| 1 | abc | 1 | 3 | ... |
| 2 | acb | 0 | 2 | ... |
| 3 | bca | 3 | 2 | ... |

TABLE II
SAMPLE TABLE FOR CHEMISTRYMARKS

C. Query analysis

Query 1: select * from physicsmarks where m1 = 1
 This was a very simple query and the threshold was crossed. After JIT Indexing was done the query resource consumption was brought down to one third of the original resource consumption. The cost of index creation is equal to the cost without JIT which is nothing but the cost for a complete table look up. If such a query is repeated three times, the cost of index creation can be justified.

Query 2: select * from physicsmarks where m1 =2 and m2 = 1
 This is a bit more resource consuming query and the threshold was crossed. After JIT Indexing was done the query resource consumption was brought down considerably.

Query 3: select * from physicsmarks where m1 = m2
 In this query the JIT Indexing scheme failed. Even though a Just In Time Index was created, the index turned to be of no use and the query was answered using the conventional table look up methods. More intelligence has to be provided for the JIT Indexer process for avoiding such kind of situations.

Query 4: select * from chemistrymarks INNER JOIN physicsmarks on chemistrymarks.m1 = physicsmarks.m1
 This was a very much resource intensive query and it needed a join on two tables for answering the query. The real application of Just In Time Index-ing is demonstrated with this sample query. After creating the index, the resource consumption for answering the query is brought down gigantically. The cost of index creation is nothing but the cost for a complete table look up. From the second time onwards this query is going to be answered in a flash with the help of JIT Index.

| Cost without JIT | 2340508.672 |
|---|---|
| Initial Scan Object | Table |
| Threshold | 50 |
| Index created by JIT | index(m1) |
| Cost with JIT Index | 749849.5149687501 |
| Final Scan Object | Index |

TABLE III
ANALYSIS FOR QUERY 1

| Cost without JIT | 2340508.672 |
|---|---|
| Initial Scan Object | Table |
| Threshold | 50 |
| Index created by JIT | index(m1) |
| Cost with JIT Index | 2340508.672 |
| Final Scan Object | Table |

TABLE V
ANALYSIS FOR QUERY 3

| Cost without JIT | 2340508.672 |
|---|---|
| Initial Scan Object | Table |
| Threshold | 50 |
| Index created by JIT | index(m1,m2) |
| Cost with JIT Index | 148706.27207788086 |
| Final Scan Object | Index |

TABLE IV
ANALYSIS FOR QUERY 2

| Cost without JIT | 233691.70799999998 |
|---|---|
| Initial Scan Object | Table |
| Threshold | 50 |
| Index created by JIT | index(physicsmarks(m1), chemistrymarks(m1)) |
| Cost with JIT Index | 7.403727970634882E10 |
| Final Scan Object | Index |

TABLE VI
ANALYSIS FOR QUERY 4

## X. CONCLUSION AND FUTURE WORK

The achieved results clearly support the scope of the proposal of Just In Time Indexing. One of the major challenges being faced by Database managers today is to manage the performance of complex SQL queries which are dynamic in nature. Since it is not possible to tune each and every query because of its dynamic nature, there is a definite need of an automated system which can monitor each and every query and decide whether any temporary index can do the job better or not. Also the maintenance of such temporary indexes is going to be a challenge since there are many factors affecting that. The resource consumption calculation and estimation of threshold are the real problems with the implementation of this algorithm. It is because for resource calculation of queries one need to get into the lower level details of implementation of the database manager and one need to have a good understanding of the query execution plan.

The bringing concept of multiprocessor system into this algorithm has its on pros and cons. Efficiency will be more, but to switch the query execution from one plan to another in the middle of execution is going to be a major challenge. The advancements so far achieved shows that this project can be implemented in a full fledged manner.

Since this is the era of multi-processor systems, the concept of parallelism can be brought down to JIT Indexing also. The key idea is to, while the JIT Indexing engine will be making an index, the other processors or cores can be used to build the result set using an existing dense index or by using the relation itself. While the JIT Indexing engine is ready with the temporary sparse index, processing of the query using the conventional method is stopped and a live switch over can be made to the temporary JIT Index so that the rest of the processing can be made using this temporary index. The processing of the query using the JIT Index will start from the point where the first processing was stopped using the conventional method.

For this additional feature to be implemented, several key issues are to be addressed. The decision problems like

1) Is this machine a multi processor system?
2) Is this query is going to have a running time

| ID | Cost without JITI | Scan Object | Threshold |
|---|---|---|---|
| 1 | 2340508.672 | Table | 50 |
| 2 | 2340508.672 | Table | 50 |
| 3 | 2340508.672 | Table | 50 |
| 4 | 233691.708 | Table | 50 |

TABLE VII
RESULTS IN A GLANCE DURING JIT ALERT PROCESS

| ID | Index on | Cost with JITI | Scan Object |
|---|---|---|---|
| 1 | m1 | 749849.515 | Index |
| 2 | m1,m2 | 148706.27 | Index |
| 3 | m1 | 2340508.672 | Table |
| 4 | m1,m2 | 7.404 | Index |

TABLE VIII
RESULTS IN A GLANCE DURING JIT INDEXER PROCESS

which is greater than a threshold?
3) Will the query can perform better with the help of an index?

The answers for these questions will be the deciding factor for making the decision about temporary index if the whole system is implemented in a multi processor system.

There can be other problems like while the temporary sparse index is being made up, the task might have completed or got killed. In such situations the temporary index has to be either discarded or maintained for future use. This idea can be used in situations where it is not clear whether using an existing index or creating a temporary index will make the greatest query efficiency. The query optimiser starts running the query using an existing index, while in the background and in parallel it will build a temporary index for the query.

REFERENCES

[1] Postgresql 9.1.2 documentation: Create table.
[2] Elena Barcucci, R. Pinzani, and Renzo Sprugnoli. Optimal selection of secondary indexes. IEEE Trans. Softw. Eng., 16(1):32–38, January 1990.
[3] Douglas Comer. The difficulty of optimum index selection. ACM Trans. Database Syst., 3(4):440–445, December 1978.
[4] Douglas Comer. The ubiquitous b-tree. Computing Surveys, 11(2):123–137, 1979.
[5] Paul Reuben Day. Method, query optimizer, and computer program product for implemeting live switchover to tem- porary sparse index for faster query performance, 11 2004.
[6] B.-J. Falkowski. Comments on an optimal set of indices for a relational database. Software Engineering, IEEE Transactions on, 18(2):168 –171, feb. 1992.
[7] S. Finkelstein, M. Schkolnick, and P. Tiberio. Physical database design for relational databases. ACM Trans. Database Syst., 13(1):91–128, March 1988.
[8] Johann Christoph Freytag. Query Processing for Advanced Database Systems. Morgan Kauffmann Publishers, 1994.
[9] Michael Hammer and Arvola Chan. Index selection in a self-adaptive data base management system. In Proceedings of the 1976 ACM SIGMOD international conference on Management of data, SIGMOD '76, pages 1–8, New York, NY, USA, 1976. ACM.
[10] Jennifer D. Widom Hector Garcia-Molina, Jeffrey D. Ullman. Database Systems: The Complete Book.
[11] M. Y. L. Ip, L. V. Saxton, and V. V. Raghavan. On the selection of an optimal set of indexes. IEEE Trans. Softw. Eng., 9(2):135–143, March 1983.
[12] Raghu Ramakrishnan and Johannes Gehrke. Database Management Systems. Mc-Graw Hill International Editions, 2nd edition, 1999.
[13] Dennis Shaha and Philippe Bonnet. Database Tuning Principles, Experiments, and Troubleshooting Techniques. Morgan Kauffmann Publishers, 2003.
[14] Art Taylor. JDBC Developer's Resource. Informix Press, 1997.

[]